\documentclass[a4paper,12pt]{article}
\usepackage{graphicx}
\usepackage{natbib}         
\usepackage{subfig}

\newcommand{\arcsec}{$^{\prime\prime}$}

\chardef\us=`\_

\usepackage{psfig}          
\begin{document}

\begin{center}
{\bf \Large
Differential Emission Measure Evolution as a Precursor of Solar Flares
}\\
{\bf C.~Gontikakis$^{1}$ 
        I.~Kontogiannis$^{2}$
        M.K.~Georgoulis$^{1, 3}$
        C. Guennou$^{4}$
P. Syntelis$^{5}$\,
S.H. Park$^{6}$\, E.~Buchlin$^7$ }
\end{center}
\begin{center}
$^{1}$ Research Center for Astronomy and Applied Mathematics, Academy of Athens,\\ Soranou Efesiou 4, 11527, Athens,\\                     
              $^{2}$ Leibniz-Institut fur Astrophysik Potsdam, Germany \\
             $^{3}$ Department of Physics \& Astronomy, Georgia State University, Atlanta, GA 30303, USA \\ 
             $^{4}$ Talan, Sacrebleu LLC, New York, USA \\ 
	      $^{5}$  School of Mathematics and Statistics, St. Andrews, United Kingdom \\
              $^{6}$ Institute for Space Earth Environmental Research, Nagoya University, Nagoya, Japan \\ 
              $^7$ Institut d'Astrophysique Spatiale, CNRS, Univ. Paris-Sud, Universit\'e Paris-Saclay, B\^at. 121, F-91405 Orsay cedex, France

\date{Received / Accepted}
\end{center}
\begin{abstract}
In this work we analyse the temporal evolution of the Differential Emission Measure (DEM) of solar active regions and explore its usage and credibility as a predictor or precursor of major solar flares. 
We use full disk DEM maps, provided by the Gaussian Atmospheric Imaging Assembly (GAIA-DEM) maps archive. 
These are calculated using the Extreme UltraViolet (EUV) images by the {\it Atmospheric Imaging Assembly} (AIA) imager, on-board the {\it Solar Dynamic Observatory} (SDO), assuming a Gaussian dependence of the DEM on the logarithmic temperature. 
Our analysis is based on two datasets, namely, one including time-series of sixteen (16) solar active regions and another including a statistically significant sample of 9454 point-in-time observations corresponding to hundreds of NOAA-labelled and unlabelled regions observed during solar cycle 24. From the temporal evolution of the DEM-related parameters, we find that the temporal derivatives of two parameters, namely, Emission Measure $d{\rm EM}/dt$ and maximum DEM temperature $dT_{\rm max}/dt$ frequently exhibit high positive values a few hours before M- and X-class flares, indicating that flaring regions become brighter and hotter as the flare onset approaches. From the point-in-time observations we compute the conditional probabilities of flare occurrences using the distributions of positive values of the $d{\rm EM}/dt$, and $dT_{\rm max}/dt$ time series and compare them with corresponding flaring probabilities of the total unsigned magnetic flux of active regions, that is a conventionally used, standard flare predictor. For C-class flares, conditional probabilities have lower or similar values with the ones derived using the unsigned magnetic flux for 24 and 12hours forecast windows, while for M-class and X-class flares, these conditional probabilities are higher than those of the unsigned flux for higher parameter values.
Shorter forecast windows of the order of 2-12 hours improve the conditional probabilities of $d{\rm EM}/dt$, and $dT_{\rm max}/dt$ in comparison to those of the unsigned magnetic flux. We conclude that flare forerunner events such as preflare heating or small-flare activity prior to relatively major flares (of M-class and above) reflect on the temporal evolution of EM and $T_{\rm max}$. Of these two, high values of the temporal derivative of the EM could conceivably be used as a credible precursor, or short-term predictor, of an imminent flare.
\end{abstract}
keywords: Flares, Forecasting;Flares, Pre-Flare Phenomena; Spectrum, Ultraviolet

\section{Introduction}
Solar flares consist of explosive releases of magnetic energy, accompanied by enhanced emission throughout the electromagnetic spectrum see \cite{2008LRSP....5....1B} for a review. 
Flares are classified in A, B, C, M, X classes, according to the logarithm of their 1 - 8~\AA\ soft X-ray peak flux (W m$^{-2}$), as observed by GOES satellites. In this study we will focus on
two groups of flares; those between $10^{-6}$ and $10^{-5}$ W m$^{-2}$ (C-class) and those above $10^{-5}$ W m$^{-2}$ (M-class and X-class). We will refer to the latter  flares as \lq major flares\rq . 

Regarding flare prediction, morphological physical quantities derived from measurements of active-region photospheric magnetic fields, seem to hold more promise and, so far, the study of precursor signals has relied mostly (but not exclusively) on photospheric vector or line-of-sight (LOS) magnetograms  \citep[see e.g.][]{2013Entrp..15.5022G}. However, it has been long known that flare precursors include phenomena that take place in various heights in the chromosphere and corona of active-regions \citep{1980SoPh...68..217M}.

Most recent observations have systematically shown the existence of precursor signals in EUV and X-ray observations \citep{2011SSRv..159...19F}, but only recently these have been utilized in solar flare forecasting \citep{2017ApJ...835..156N,2018SoPh..293...48J}. Precursor signals can be small X-ray bursts \citep{1991A&AS...87..277T, 1998SoPh..183..339F} or pre-heating events \citep{1987SoPh..107..271L} most of which happen a few (up to 40) minutes before the flare. Such a pre-heating phase was detected in soft X-rays, a few minutes before the onset of the flare's impulsive phase by \citet{2002SoPh..208..297V}. Similarly, using EUV spectroscopy, \citet{2009ApJ...691L..99H}, found an increase of the non-thermal witdth of the Fe xii/xxiv 195.1\AA\ spectral line, a few hours prior to an X-class flare. This increase was found to follow helicity injection measured by the photospheric magnetic field. More recently, \citet{2016MNRAS.459.3532G}, studied potential precursors of flares in X-rays, taking place as early as 24 hours before the main flare onset.

Plasma heating is often probed through
the differential emission measure (DEM). The DEM, shows the distribution of the plasma Emission Measure (EM) as a function of temperature and quantifies the temperature structure of the optically thin coronal and transition region plasma, along the observational line of sight (LOS). The emission measure is defined as ${\rm EM}\,=\, n_e^2 L$, where $n_e$ is the electron density and $L$ is the length of the emitting plasma along the LOS. 
The DEM has been frequently used as a diagnostic of thermal conduction and plasma heating \citep{2011ApJ...740....2W, 2013ApJ...774...31G,2015Ge&Ae..55..995M}, of thermal processes in pulsating loops \citep{2015ApJ...807..158F} and it can also be used to characterize the morphology and temporal evolution of solar eruptions \citep{2012ApJ...761...62C,2013A&A...553A..10H,2014ApJ...786...73S}.

\citet{2016A&A...588A..16S} used DEM analysis to study the preflare heating in AR 11429. They found an almost steady four-hour-long increase prior to two X-class flares  associated with the launch of two Coronal Mass Ejections (CMEs). This was interpreted as the signature of heating taking place in the rising pre-CME flux rope, enhancing the 7~MK hot plasma component before the launching of the two erupting X-class flares.

The Flare Likelihood and Region Eruption foreCASTing (FLARECAST, www.flarecast.eu) project provided a solar flare prediction facility that can be either standalone or integrated under wider and more general space weather forecasting schemes in the future. 
 During this joint effort, established flare predictors were adapted and the efficiency of new ones was investigated, leading to a variety of properties that characterize the eruptive potential of active regions 
\citep[see e.g.][]{Kontogiannis17,Kontogiannis18,Guerra18,Park18}. 

In this context, here we explore the feasibility of using EUV observations as precursors of flaring activity. To this end, we assess the statistical significance of the results of \citet{2016A&A...588A..16S} by analysing large samples of active regions and explore the efficiency of DEM-related parameters as flare predictors. The paper is organized as follows: In Section 2 we present the datasets and our method. In Section 3 we investigate time-series of DEM-related parameters and their characteristics in respect with flare occurences. In Section 4 we use a statistically significant sample to assess the potential of DEM-related parameters as flare predictors. Our results are discussed in Section 5.

\section{Observations and Data Treatment}
For this study we utilize data provided by two databases. The first one is the Gaussian Atmospheric Imaging Assembly Differential Emission Measure (GAIA-DEM) maps archive, provided by the {\it Multi Experiment Data Operation Center} (MEDOC) data center at Institut d'Astrophysique Spatiale
(IAS) in Orsay. The second database contains photospheric vector magnetograms taken by the {\it Helioseismic and Magnetic Imager}
\citep[HMI][]{2012SoPh..275..207S,2012SoPh..275..229S} onboard the {\it Solar Dynamic Observatory} \citep{2012SoPh..275....3P}
and more specifically the Space Weather HMI Active Region Patches (SHARP \citet{2014SoPh..289.3549B}).

To explore the association of DEM-related parameters we create two data sets from the databases: the first is a small sample of active-region time-series where we verify our methodology aimed to assess whether the DEM parameters have the potential to be
used as precursors of solar flares. The second dataset is an extensive
statistical sample of point-in-time observations, consisting of more than 9000 cut-outs of regions, where we apply our methodology and examine whether the DEM parameters can be used as predictors in a realistic flare-prediction scenario.

\subsection{The GAIA-DEM archive}
For a number ${\rm N}$ of spectral lines or filtergrams, the DEM is defined by N integrals of the form $I_i\,=\,\int DEM(T) G_i(T)dT$ where $I_{\rm i}$, ${\rm i=1,..N}$ are the spectral line or filtergram intensities and $G_i(T)$ are the corresponding contribution functions of the temperature domain. The contribution functions contain the emission terms and the instrument response parameters, as obtained by the atomic characteristics of the emission lines or filtergrams.

For the DEM distribution, the GAIA-DEM archive utilizes the six intensities of the coronal EUV channels, (94\AA, 131\AA, 171\AA, 193\AA, 211\AA, and 335\AA) observed with the AIA imager \citep{2012SoPh..275...17L} on board SDO. For each filter, 10 successive images, with a one minute cadence, are summed, and rebinned to a spatial resolution of 2.4\arcsec, in order to increase the signal-to-noise ratio.
On the simplified assumption that the DEM of each pixel in the FOV follows a Gaussian distribution, four parameters are calculated through a fitting procedure. These parameters are the EM (i.e. the the temperature integral over the DEM, in units cm$^{-5}$), the temperature corresponding to the position of the Gaussian function maximum 
($T_{\rm max}$, in Kelvin), its width $\sigma$, (in $\log_{10}(T)$) and the $\chi^2$ of the fit procedure. The Gaussian representation of the DEM distribution is considered successful, when $\chi^2 <4$  \citep{2012ApJS..203...25G, 2012ApJS..203...26G}.

There is no \textit{a priori} physical reason for the DEM distribution to exhibit a Gaussian temperature function. There are methods that provide more complex DEM functions. However, this working hypothesis gives a first description of the DEM through the three parameters.

Starting from 13 May 2010, the GAIA-DEM archive provides the corresponding full-disk maps for each parameter.  These maps are provided every 30 minutes or 1 hour, depending on the date of the observation, (see http://medoc-dem.ias.u-psud.fr/sitools/client-user/DEM/project-index.html for more details on the DEM map computation).

\subsection{Active Region timeseries}
Table~\ref{firstset} contains the sixteen National Oceanic and Atmosphere Administration (NOAA) active regions used for this study, along with information on their duration of observation as they traverse the solar disk, their start and end heliographic longitudes, and their flare productivity. Our sample comprises the eleven active regions studied by \citet{Kontogiannis17} (see their Table~1), to which we added five non-flaring active regions to enhance the statistics of the non-flaring regions.

The flare association, i.e. the magnitudes of flares that took place at each selected active region during the studied time and their onset times, was performed using the FLARECAST and the Geostationary Operational Environmental Satellite (GOES) databases and the Solar Monitor website (https://www.solarmonitor.org).

From the full-disk maps we cropped rectangle fields-of-view (FOV) that contained each active region, avoiding any bright EUV structures from neighbouring regions. To track the active regions throughout the corresponding observing period, the full-disk time-series were corrected for differential rotation using the formula of \citet{1990ApJ...351..309S}. In some cases, this rotation formula was not sufficient because active regions can have different rotation rates depending on their age and on the size of their sunspots \citep{2000SoPh..191...47B}. In these cases we adjusted the correction by means of visual inspection. As an example, Figure~\ref{demmap} shows the cropped regions for NOAA AR 11428 and 11429 on the four maps of 6 June 2012 at 23:05:14 UT.

\begin{table}
\centering
\begin{tabular}{llllrrr}
NOAA  & start time&end time&longitudes [$^{\circ}$]&X&M&C \\ \hline
11429$^{\star}$&2012-03-04 00:05&2012-03-10 23:05& -73\hspace{0.5 cm}26  &3      & 13 & 37\\
11875&2013-10-18 00:04&2013-10-28 23:04& -88\hspace{0.5 cm}64  &1      & 9 &  56\\
11158&2011-02-10 22:04&2011-02-15 23:05& -40\hspace{0.5 cm}23  &1      & 1  & 12\\ 
11748&2013-05-15 00:04&2013-05-18 23:05& -64\hspace{0.5 cm}23  &1      & 2  &  7\\
11515$^{\star}$&2012-06-28 00:05&2012-07-06 23:35& -84\hspace{0.5 cm}42  &0      & 18 &  60\\ 
11513$^{\star}$&2012-06-28 00:05&2012-07-06 23:35& -48\hspace{0.5 cm}68  &0      & 7  & 22 \\
11428$^{\star}$&2012-03-04 00:05&2012-03-10 23:05& -44\hspace{0.5 cm}46  &0      & 0  & 4 \\
11640&2013-01-01 00:05&2013-01-05 23:05& -06\hspace{0.5 cm}69  &0      & 0  & 2 \\
11663&2013-01-29 00:05&2013-02-03 23:05& -20\hspace{0.5 cm}65  &0      & 0  & 1 \\
11072&2010-05-21 17:35&2010-05-24 23:04& -22\hspace{0.5 cm}24  &0      & 0  & 0 \\
11156&2011-02-10 22:04&2011-02-15 23:05& -52\hspace{0.5 cm}15  &0      & 0  & 0 \\
11157&2011-02-10 22:04&2011-02-15 23:05& -53\hspace{0.5 cm}15  &0      & 0  & 0 \\
11638&2013-01-01 00:05&2013-01-05 23:05& -24\hspace{0.5 cm}38  &0      & 0  & 0 \\
11639&2013-01-01 00:05&2013-01-05 23:05& -11\hspace{0.5 cm}58  &0      & 0  & 0 \\
11863&2013-10-10 00:04&2013-10-13 23:04& -07\hspace{0.5 cm}48  &0      & 0  & 0 \\
11923$^{\star}$&2013-12-12 00:05&2013-12-15 23:35& -21\hspace{0.5 cm}39  &0      & 0  & 0 \\
\end{tabular}
\caption{Active regions and flares of the first data set. Stars indicate active regions sampled with 30~min cadence. 
Absence of stars indicate 1~hr cadence. NOAA 11429 and 11428 have a 30 min cadence during 6 days.}
\label{firstset}
\end{table}

Then, we calculated the average value of each of the DEM parameters inside these FOVs, excluding the parameter values for which
$\chi^2 \geq\ 4$. This process resulted in four timeseries per active region, describing the time evolution of the DEM parameters. We performed interpolations so that all time-series have a time step of 0.5~hours. The behaviour of the timeseries was analysed by computing their time derivatives i.e.,
$d{\rm EM}/dt$, $dT_{\rm max}/dt$, $d\sigma/dt$  and $d\chi^2/dt$. We calculated the derivatives using a Savitzky-Golay filter (SG) \citep{1964AnaCh..36.1627S} on selected sections of the timeseries. The SG filter assumes polynomials to connect adjacent data points, and can be used to smooth a timeseries similarly to any polynomial interpolation smoothing function. The coefficients of the
fitted polynomials can be used to calculate the derivatives of the timeseries using the idl convol routine. We apply the SG filter to smooth out the small variations of the timeseries, to bring out the possible preheating time variation, and to calculate the derivatives of the smoothed signal. After some testing, we selected a SG filter that smooths the data with a 3-degree polynomial  computed on each point $t_{\rm i}$ of the timeseries by taking into account its closest points, from $t_{\rm i-3}$ to $t_{\rm i+3}$.
The calculation of the derivatives was also performed with the common central-difference approach where the derivative of a parameter at time $t{\rm _i}$ is obtained by the half-difference of parameter values at times $t_{\rm i-1}$ and $t_{\rm i+1}$ in order to check if we found significant variations between the two methods.
Then we calculated the derivatives of 5 to 5.5hours long sections of the time-series that do not include the flares under study, to ensure that the flare peak values are not involved in the SG calculation. The results of both approaches will be compared in the next sections.

\subsection{Point-in-time observations}
For the second set of GAIA-DEM maps we use as a reference a representative sample of SHARP data of Solar Cycle 24. SHARPs are cut-outs of regions of interest such as active regions and plages and contain maps of the magnetic field components, the Doppler velocity and the continuum intensity. They also include a set
of parameters intended as predictors. We utilized the Near-Real-Time (NRT) Cylindrical Equal
Area (CEA) version of SHARP. Our sample was constructed as follows: for 336 randomly selected days between
September 2012 and May 2016, we selected all data with a 6\,h cadence, with no restrictions on the position of the active region on the solar disk. This resulted in a total of 9454 SHARP cut-outs, out of which 3087 correspond to 490 NOAA active regions, while the remaining 6367 correspond to active regions with no NOAA identifier.

The corresponding GAIA-DEM cut-outs were taken at the closest earlier time, $t_0$, of the SHARP image time, $t_{\rm SHARP}$. Each region was then tracked for the 6 hours that preceded $t_0$, using again the treatment for differential solar rotation described in the previous subsection. Since the DEM map cadence is either 30 minutes or one hour and it may also contain gaps, the number of DEM measurements for each point-in-time observation may be 12, 6 or fewer. 
\begin{table}
\begin{center}
\centering
\begin{tabular}{rrrll}
after $t0$[hr] & X  & M & C  &no flares\\  \hline
2              & 2  & 14 &202 &  8868  \\
6              & 4  & 30 &395 &  8677  \\
12             & 7   & 67  & 652& 8417 \\
24             & 12 &125& 976  &  8093 \\
\end{tabular}
\caption{Number of flares for each GOES magnitude class, for different time intervals after the end times of the 9083 SHARP image timeseries.}
\label{secondset}
\end{center}
\end{table}
For each of these 6-hour long time series we followed the same methodology described above, i.e. we excluded the values of the FOV for which $\chi^2 \geq\ 4$, we calculated the temporal averages of each parameter to create 6-hour long time-series of the four DEM parameters. We interpolated the timeseries in order to have a 30-minute time step in all cases, we deleted the ones having data gaps and we determined the corresponding derivatives. Then we counted the number of flares that occurred within time windows of 24, 12, 6 and 2 hours from the acquisition of each of the remaining 9083 cut-outs (Table~\ref{secondset}).  The efficiency of the of DEM-related parameters as predictors/precursors will be compared with that of the total unsigned magnetic flux, $\Phi$, which was calculated for each of the SHARP cut-outs of our sample. $\Phi$ is considered is a rudimentary quantity that may serve as a reference flare predictor \citep{2003ApJ...595.1296L, 2012SoPh..276..161G}. For both datasets, the two major flare classes (M- and X-class) have been treated jointly due to the relatively poor statistics caused by the small number of X-class flares of Solar Cycle 24.

\subsection{Parameter uncertainties}

The typical uncertainties of the DEM parameters $T_{\rm max}$ and $\sigma$ can be estimated, using the calculations performed by \citet{2012ApJS..203...25G,2012ApJS..203...26G}. In Figure 6e,f,g of \citet{2012ApJS..203...25G} it is shown that a typical value of the uncertainty is $\Delta T_{\rm max}\simeq\ \pm 100\,000$~K when $T_{\rm max}=2.\times\ 10^6$~K,  which corresponds to a relative error of 10\%. In Figure 10g of \citet{2012ApJS..203...25G}, the relative error $\Delta \sigma / \sigma$,  can be of the order 150\% when $\sigma=0.2$ and is increasing for higher $\sigma$ values. 
The uncertainties of the EM are not included in \citet{2012ApJS..203...25G,2012ApJS..203...26G}. Since the EM and $T_{\rm max}$ represent the zero and first order moments of the DEM distribution, correspondingly, it is reasonable to assume that the relative error of the EM is, at most, of the same order of the relative error of $T_{\rm max}$.

For each point of our timeseries we average N values included in each FOV. The averaged $T_{\rm max}$ values are associated with a typical error of $\Delta T_{max}/\sqrt{N}$. For the data first dataset, N $\in$ (7500, 13000) while for the second N $\in$ (1000, 5000).

In \citet{2012ApJS..203...25G, 2012ApJS..203...26G} the uncertainty calculations take into account systematic errors that we ignored as they do not influence the timeseries behavior. 
Therefore, the uncertainties of the EM and $T_{\rm max}$ timeseries are expected to decrease significantly due to averaging, resulting in timeseries of averaged variables with small relative errors.

\section{Analysis of the first data set}

\begin{figure}
        \centering
\includegraphics[width=1.\columnwidth]{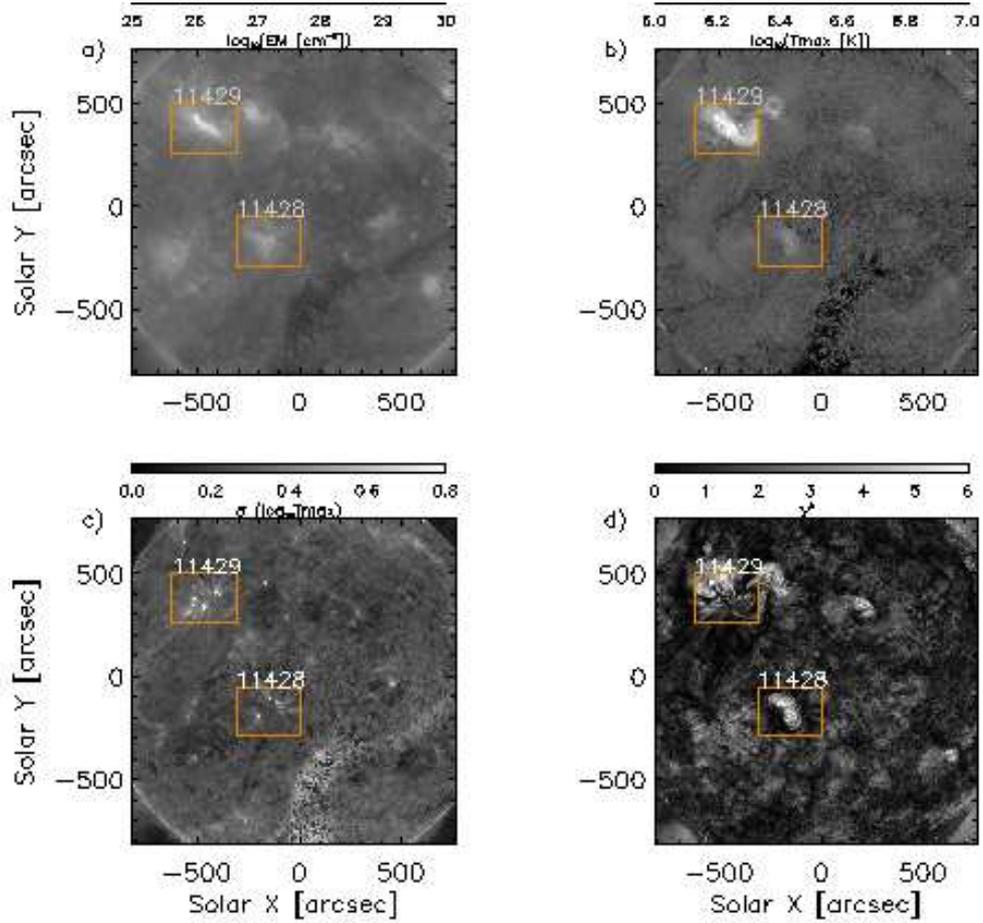}
\caption{GAIA-DEM parameter maps of 6 June 2012 at 23:05:14~UT. The green rectangles show the sub-FOV for the NOAA ARs 11429 and 11428. EM (panel a), maximum temperature $T_{\rm max}$ (panel b), DEM width $\sigma$ (panel c) and $\chi^2$ of the fit (panel d).}\label{demmap}
\end{figure} 
\subsection{Detailed analysis of two active regions}
We first focus on two of the active regions of the first dataset to describe in more detail how we derive timeseries from the DEM maps.

Figure~\ref{demmap} shows the full-disk maps of the four DEM parameters on 6 June 2012, when NOAA active regions 11428 and 11429 are visible.

The two active regions are seen clearly in the EM and in the maximum temperature $T_{\rm max}$ (Figure~\ref{demmap}a,b). The higher EM is caused by the higher EUV emission of the active regions 
compared with the quiet Sun. NOAA 11429 includes a hot flux rope, surrounded by high temperature loops \citep{2015ApJ...809...34C, 2016A&A...588A..16S}. This 
flux rope was destabilized producing two X-class flares.

The DEM width map (Figure~\ref{demmap}c), shows mostly the legs of loops anchored at the periphery of the active regions,  whereas hot active region cores are showing smaller widths. Higher widths at the active region periphery may be linked to a cool plasma component that results in a wider temperature range in the DEM distribution. Finally, as seen in Figure~\ref{demmap}d, within the sub-FOV, many  pixels have $\chi^2$ values that are higher than 4 and, as mentioned in Sect.~2.2, these are not considered for the calculation of temporal averages.

\begin{figure}
        \centering
\includegraphics[width=.95\columnwidth]{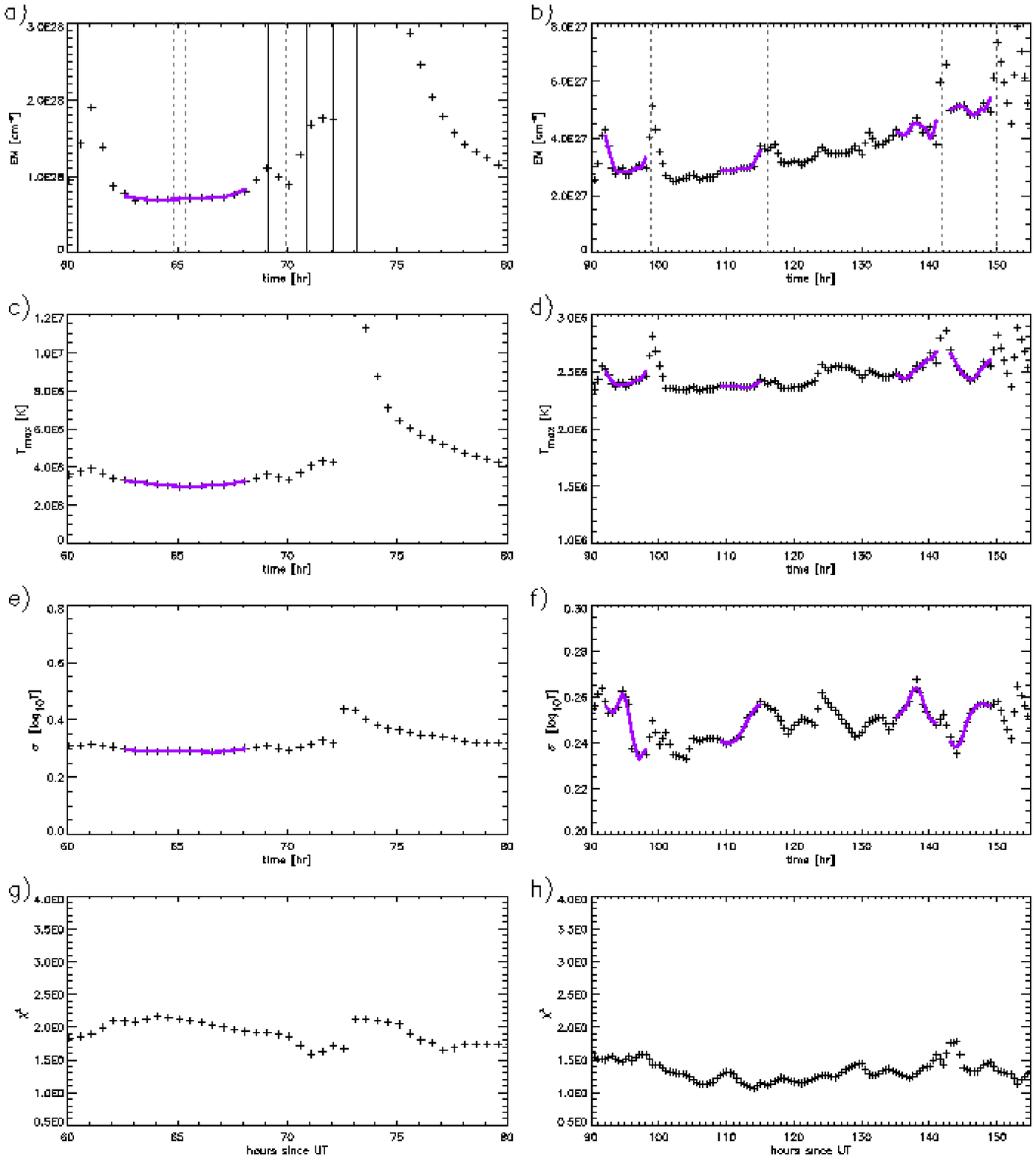}
\caption{Parts of timeseries of DEM parameters from active regions NOAA 11429 (first column) and NOAA 11428 (second column). First, second, third, and fourth lines show EM, (in cm$^{-5}$), $T_{\rm max}$ (in K), $\sigma$ (in $\log_{10}$ [K ]) and $\chi^2$ respectively. The crosses show the data points and the thick purple lines a Savitzky-Golay filter over the 6hr sections of the data selected to study preheating events. The NOAA 11429 timeseries segment starting time is 2012-03-06, 10:05~UT and the NOAA 11663 is 2012-03-06 01:35 UT as indicated below panels g and h.}\label{timeserie_11429_28} 
\end{figure}

Figure~\ref{timeserie_11429_28} shows segments of the timeseries of the averaged DEM parameters for both active regions.
The original timeseries (black crosses) do not differ significantly from the ones smoothed with the SG filter calculated over 6hr segments. Although smoothing results in lower peaks, these time-series still capture all the important morphological characteristics of the DEM parameter evolution.
Flare onsets coincide with distinct peaks in the EM, $T_{\rm max}$ and $\sigma$ timeseries
(Figure~\ref{timeserie_11429_28}a,b,c) which are more prominent for the two X-class flares of NOAA AR 11429.
These flares, with magnitudes X5.4 and X1.3 are the largest flares of our sample.
The other M-class and C-class flares show a much smaller trace in the timeseries. Notice also the increase exhibited by
the values of EM, $T_{\rm max}$ and $\sigma$, before the X-class flares, that could be interpreted as a hours-long pre-heating, while after the flare peaks, the values decrease again 
gradually due to plasma cooling. On the contrary, $\chi^2$, does not exhibit any notable response to pre-flare heating or flaring   (Figure~\ref{timeserie_11429_28}d).

The thick purple  smoothed line in Fig.~\ref{timeserie_11429_28}a (between 63hr to 68hr), highlights the time interval of the pre-heating occurring in NOAA AR 11429 prior to the M1.3 flare at t=69 h. Notice that this time period is not affected from the preceding M2.1 flare ($t=60.46$~hr), and therefore shows a clear signature of the pre-heating prior to the M1.3 flare.

We can compare the timeseries shown in Fig.~\ref{timeserie_11429_28}a,c with the ones presented in \citet{2016A&A...588A..16S}. There, the DEM is calculated using the \citet{2012A&A...539A.146H} method, where the functional form of the DEM has more free parameters than a Gaussian function. Therefore, one can distinguish the temporal evolution of different temperature components of the DEM. Figure~7a of \citet{2016A&A...588A..16S} shows the time evolution of the 7-12~MK section of the DEM, which 
rises by a factor of 8, during 8~hours, from 17:00~UT 6 March  2012 to 00:00~UT, 7 March 2012, just before the X5.4 flare onset that took place on 00:02~UT. In that work the DEM was integrated over the small area of the active region where the X5.4 class flare later took place..

In comparison, in Fig.~\ref{timeserie_11429_28}a,c from 63hr to 67hr, the EM timeseries rises by 10\%  and the $T_{\rm max}$ one by a factor 3\%. This is a small but continuous rise during 4 hours. Moreover, because we average over a few thousand pixels, the error bars are or the order of 1/1000 and so this result is meaningful. 
 Our calculations are less sensitive than the ones in \citet{2016A&A...588A..16S}, because of the Gaussian assumption of the DEM distribution, and because we average over the entire active region FOV.  However, our approach is also able to detect the  increase of the DEM parameters as a function of time, similarly to \citet{2016A&A...588A..16S}.

NOAA AR 11428,  exhibits notable differences from NOAA AR 11429. This active region is far less flare-productive (four C-class flares, see Table~\ref{firstset}). Both EM and $T_{\rm max}$ (Fig.~\ref{timeserie_11429_28}b,d) are one order of magnitude lower.
The peaks that correspond to the C-class flares are low and the EM timeseries shows a long monotonic increase (Figure~\ref{timeserie_11429_28}b).

\begin{figure}
        \centering
\includegraphics[width=1.\columnwidth]{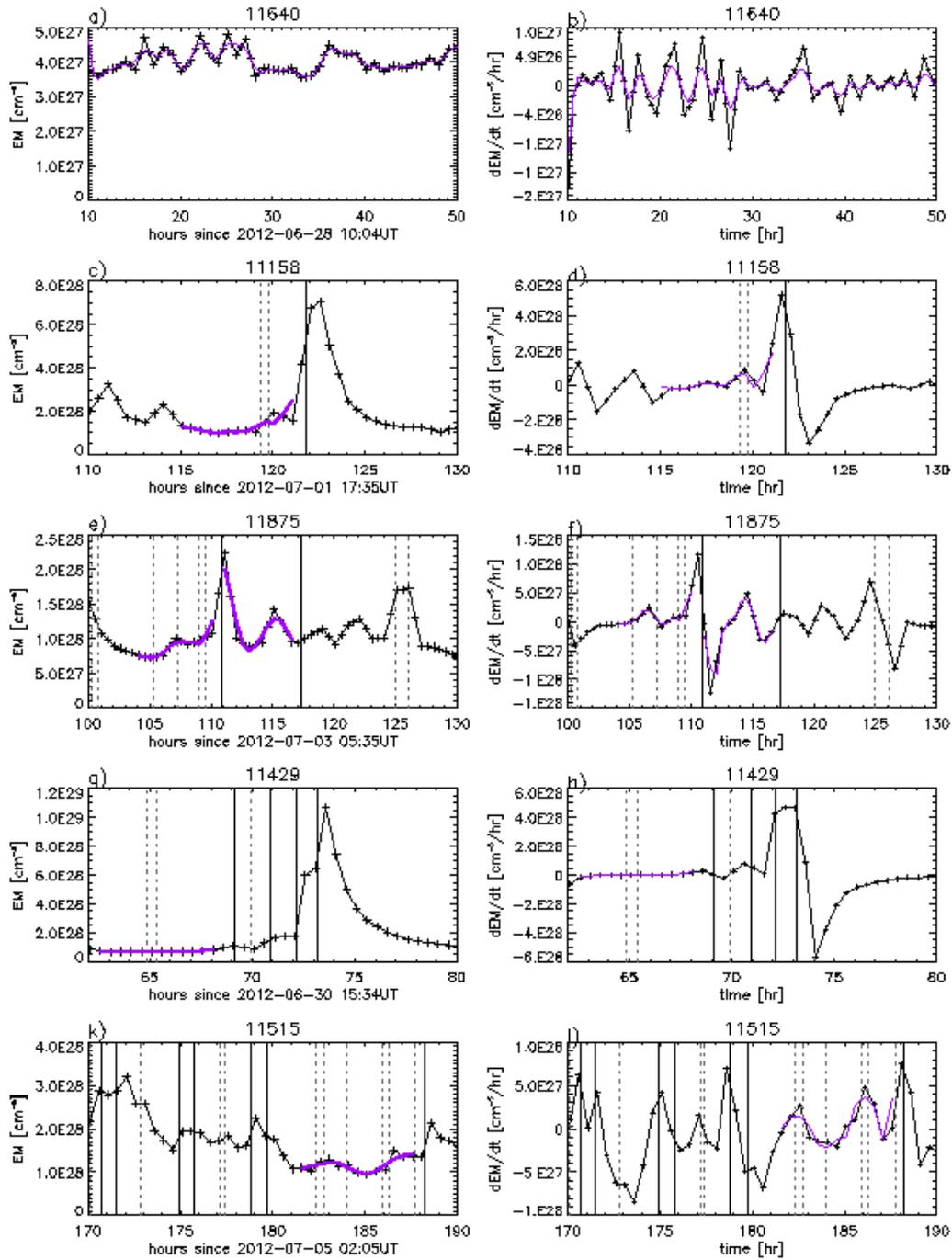}
\caption{Time sections of the EM timeseries and their derivative $d{\rm EM}/dt$ for 5 active regions from Table~\ref{firstset}.  Purple lines indicate the part of the timeseries that precedes a major (M or X-class) flare along which we sample the time derivative, except when one more major flare is preceding less than 6 hours. The onset times of major flares are indicated with vertical lines. The onset times of C-class flares are indicated with vertical short dashed lines. The blue line indicate the EM or $d{\rm EM}/dt$ timeseries where the SG filter is applied.}\label{timeserie_derivs} 
\end{figure}

\subsection{Characteristics of timeseries of the first dataset}

Having described in more detail the temporal variations of the DEM-related parameters for the two active regions in Figure~\ref{demmap}, we will show a few examples of the temporal evolution of the EM and its time derivative prior to major flares and during quiet intervals (Figure~\ref{timeserie_derivs}). We focus our discussion here on the EM but a similar analysis took place for the $T_{max}$ and $\sigma$.

NOAA AR 11640 (Fig.~\ref{timeserie_derivs}a,b) produced no flares and its EM values are roughly one order of magnitude lower than the ones exhibited by the flare-productive active regions. The temporal derivative of the EM fluctuates around zero exhibiting no notable behaviour.

NOAA AR 11158 (Fig.~\ref{timeserie_derivs}c,d) produced the first X-class flare of Solar Cycle 24. This flare was preceded by two C-class flares which seem to have resulted in increasing the EM one hour before the X-class flare. A similar behaviour is found before the M1.3 flare produced by AR 11429 (Fig.~\ref{timeserie_derivs}g,h at 69hr) and the M1.0 flare produced by AR 11875 (Fig.~\ref{timeserie_derivs}e,f 111hr).

This observation also applies to cases where more frequent and intense flaring occurs. This is the case of AR 11515 which produced repeated flaring over a few days (Fig.~\ref{timeserie_derivs}k,l).

Positive EM derivatives (\textit{i.e.} increasing EM) can also be observed before major flares during intervals with no prior flaring activity, as \textit{e.g.} in the case of the M4.2 flare of AR 11875 (at 117hr).

These observations show that there may be indications of imminent flaring activity in the temporal evolution of the EM and its derivative along certain intervals before major flaring. Within these intervals there may or may not be flares. In the following section we will examine this conjecture in a statistical manner.

\begin{table}
\centering
\begin{tabular}{lllll}
Set & No. of segments & time length & Prior to &  Flares included in timeseries\\ \hline
1   &       17        &  hours to days    &   no flares & no flares     \\
2   &       39        &    6 hours        &      C-class & no flares    \\
3   &       15        &    6 hours        &    M,X-class  &  C-class    \\
4   &       42        &    6 hours        &     M,X-class &  M,X-class  \\ \hline
\end{tabular}
\caption{Properties of four sets of timeseries segments. The segments are classified depending if they precede C-class or major flares.}
\label{statanal}
\end{table}
\subsection{Time series statistical analysis}
\label{16_ar_statistics}
Aiming to assess whether there are observable differences in the pre-flare evolution of the DEM-related parameters in respect to flaring activity, we divided the sample into four sets of 6-hour long time-segments, presented schematically in Table~\ref{statanal} and described in detail in the following:

\begin{enumerate}
\item The first set includes the timeseries from non flaring active regions, such as AR 11072. The first set also includes segments of timeseries that are flare-quiet even if they are taken from flaring active regions. These segments can be from a few hours to a few days long and are taken in active regions 11158, 11429, 11515, 11640, 11748, 11863, 11875, 11923, and 11663.

\item The second set includes the 6-hour-segments that end one hour before the onset of C-class flares. This is to ensure that the flares themselves will not affect the calculation of the derivative within these segment. We do not consider segments during which other flares of any class occurred.

\item The third set includes the 6-hour-segments that end one hour before the onset of major flares. Similarly to the second set, we excluded the cases for which major flares took place during the 6-hour segment. However, we includes cases where a C-class flare took place within the segment. One representative examples is the 6-hour-long segment before the M1.6 flare of AR 11515 in Figure~\ref{timeserie_derivs}.

\item The fourth set is similar to the third one, but this time we consider also segments during which other major flares took place. Therefore, the third set is also included in this set. 

\end{enumerate}

For these four sets of segments we examine only the positive derivatives $(d{\rm EM}/dt)^+$, $(dT_{\rm max}/dt)^+$ and $(d\sigma/dt)^+$, which can be associated with pre-flare heating. The derivatives were calculated applying the SG filter in each of the segments.

\begin{figure}
        \centering
\includegraphics[width=1.\columnwidth]{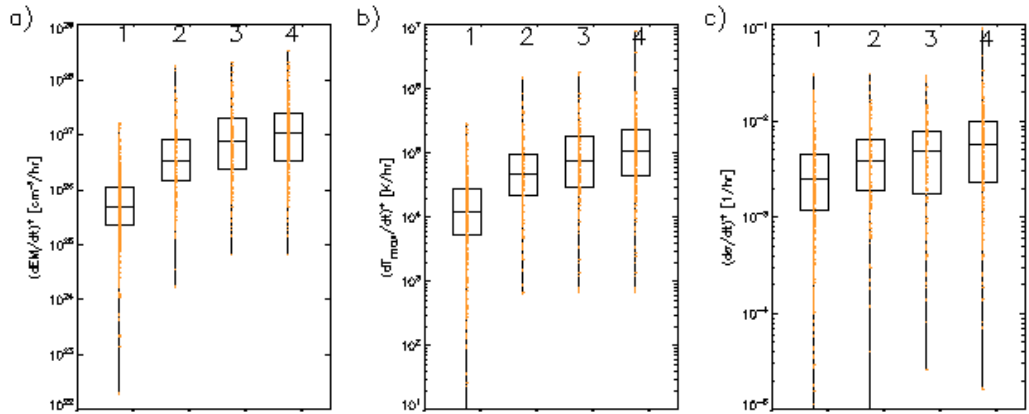}
\caption{Box-and-whisker plots calculated for the distributions of $(d{\rm EM}/dt)^+$ (a), $(dT_{\rm max}/dt)^+$ (b) and $(d\sigma/dt)^+$ (c).  For each box, the lower-to-upper horizontal lines correspond to the lower (25\%) quartile, the median and the upper (75\%) quartile of the distributions, respectively. The whiskers of the plots extend between the maximum and minimum values of the distributions, along which we also plot with colored points the data. Numbers (1)-(4) indicate the set to which each distribution corresponds (see Table~\ref{statanal}).}\label{fig_distributions}  

\end{figure}

The distributions of the values calculated for the four sets and the three DEM-related derivatives are shown in the box-and-whisker plot of Figure~\ref{fig_distributions}. The most notable feature is that, overall, the non-flaring population (\lq 1\rq\  corresponds to, on average, lower derivative values while for increasing flare class (\lq 2\rq\ to \lq 4\rq\ ) this average shifts to higher values. This indicates that there is a measurable difference in the increase of DEM-related parameters (through the higher positive values of their derivatives) prior to major flares. When we include segments during which other major flares occur before another major flare (\lq 4\rq ), these derivatives tend to increase. 
 
There is, of course, a considerable overlap of the distributions of the four sets. This overlap is greater for the $(d\sigma/dt)^+$ but for the other two derivatives, $(d{\rm EM}/dt)^+$ and $(dT_{\rm max}/dt)^+$, the 75\,\% percentiles of their distributions are clearly separated. Therefore, a $(d{\rm EM}/dt)^+$ in the range $10^{26}$ to $10^{27}$~cm$^{-5}$/hr is more likely to precede a flare than one in the range $10^{25}$ to $10^{26}$~cm$^{-5}$/hr. This is a motivation to further examine the potential of these two parameters as indicators of imminent flaring activity. 
 
More specifically, the set number 4, simulates a realistic, operational scenario of monitoring the DEM values in a real time forecast. In that case, the time-derivatives of DEM calculated during the evolution of active regions, could serve as a potential indication of future flares, even though flaring activity might take place during the examined interval. In addition to pre-flare heating, these parameters would also encompass information of the flare history, similarly to \citet{1991SoPh..131..149Z,2005PASA...22..153W,2012ApJ...757...32F}.

\section{Analysis of the point-in-time observations}

In this section we exploit the dataset of the 9083 6-hour long time series from the GAIA-DEM archive, 
which correspond to the point-in-time SHARP data, in order to assess whether the two promising DEM-related quantities, namely $(d{\rm EM}/dt)^+$, $(dT_{\rm max}/dt)^+$, could be used as flare predictors in automated forecasting schemes.

\begin{figure}
        \centering
\includegraphics[width=1\columnwidth]{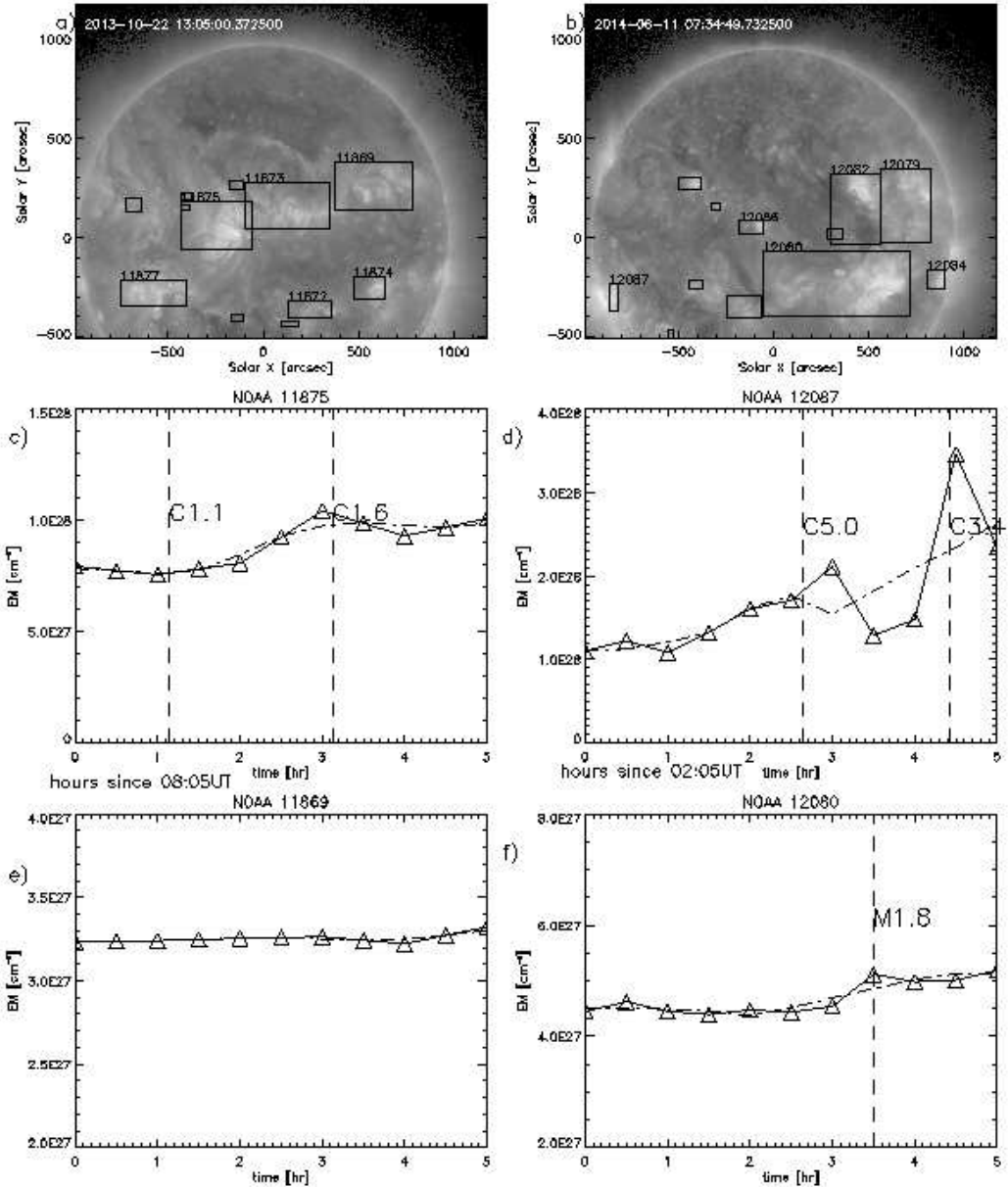}
\caption{Examples of analysis from the second data set. Panels a,b, show the EM maps for two different dates indicated in each panel. The SHARP FOVs are overplotted on the EM maps along with their corresponding NOAA numbers (if any). 
Panels c, e show the timeseries from NOAA 11875 (-245\arcsec , 61\arcsec),  and 11869 (578\arcsec , 260\arcsec\ ) 
that appear in panel a. Panel d, f show the timeseries from NOAA 12087 (-300\arcsec,-840\arcsec), and 
timeseries from NOAA 12080 (330\arcsec , -231\arcsec\ ) respectively. 
The EM timeseries are shown with triangles while the SG smoothed EM ones are shown with dashed-dotted lines. 
} \label{sharp_examples}  
\end{figure}

In Figure~\ref{sharp_examples} we present four indicative examples of flaring and quiet active regions 
taken from this dataset. Active regions 11875 and 12087 produced major flares within two hours after the end 
of the corresponding 6-hour long segments (M1.0 for the former and M3.0 followed by an X1.0 for the latter). 
Within these segments we calculate the average of the EM and $T_{max}$ derivatives, using the SG filter, 
as described in Section 2.3.  The derived values of $(d{\rm EM}/dt)^+$ will be clearly higher for the
two active regions that produce flares after the end of the segment. 
 
\begin{figure}
        \centering
\includegraphics[width=1\columnwidth]{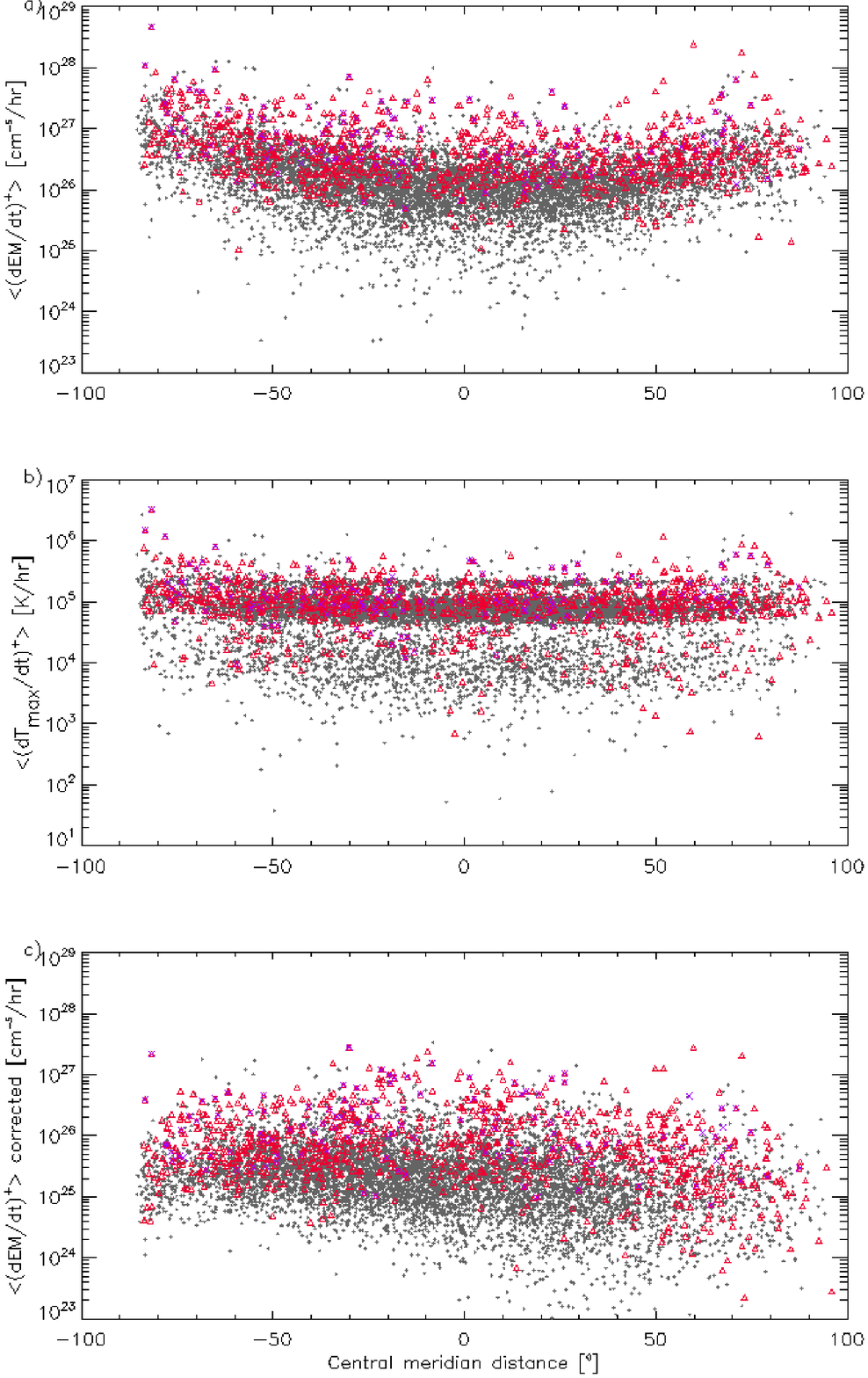}
\caption{Average positive values of the EM derivative $<(d{\rm EM}/dt)^+>$ (a) and $T_{max}$ $<(dT_{\rm max}/dt)^+>$  (b) as a function of central meridian distance. Panel (c) shows $< (d{\rm EM}/dt)^+>$ corrected for limb brightening. In all panels, \textit{red triangles} (\textit{purple crosses}) indicate cases where a C-class (M- or X-class) flare followed within 2 hours.} \label{versus_longitude}  
\end{figure}

In Figure~\ref{versus_longitude}a,b, we plot the average values of the derivatives $< (d{\rm EM}/dt)^+>$ and $<(dT_{\rm max}/dt)^+>$ calculated for each point-in-time observation, as a function of the heliographic longitude (i.e., central meridian distance). The values represented with coloured symbols show timeseries followed by flares in the next 2~hours (a C-class for the red triangles and M or X-class for the purple x-signs). We see that on average, the coloured symbols have higher values than the grey points that are not followed by flares in the next 2 hours.

As shown in Figure~\ref{versus_longitude} the values of $<(d{\rm EM}/dt)^+>$ exhibit a notable enhancement 
towards the limbs. The dependence of the EM on central meridian is a known effect, caused by the solar limb brightening 
\citep{1975SoPh...44...55M}. Towards the limb there is higher coronal emission as the line of sight crosses larger sections 
of the emitting atmosphere. Therefore, the EM timeseries show statistically lower values when regions are closer to disk center, 
as opposed to when they are closer to the solar limb. This increase is reflected on $< (d{\rm EM}/dt)^+>$ but is not detectable 
in $< (dT_{\rm max}/dt)^+>$, although hotter plasmas are expected to reach higher altitudes in the corona, 
according to their hydrostatic length scale \citep{Aschwanden_2000} and this might produce a second order center to limb 
rise in the measured $< (dT_{\rm max}/dt)^+>$.
 
We corrected the EM-based results by assuming that the coronal EUV emission comes from a spherical shell above the chromosphere. 
For each region we compute an average $\mu\,=\,cos(\theta)$ during the 6\,hours of the timeseries, where $\theta$ is the angle 
between the normal to the solar surface, and the LOS \citep{2009A&A...503..559B}. Then we multiply the EM with the corresponding 
$\mu$ value. Since EM is proportional to the section $L$ of the structure along the LOS, the quantity  $L \cos(\theta)$ approximately 
corrects for the projection effect caused by the position on the solar disk. The corrected $< (d{\rm EM}/dt)^+>$  ( Figure~\ref{versus_longitude}c) 
exhibit no clear dependence on the central meridian distance. It should be noted that the assumption of a spherical shell is a compromise, as the emission patterns in active regions originate from very complex geometries.

\begin{figure*}
        \centering
\includegraphics[width=.8\columnwidth]{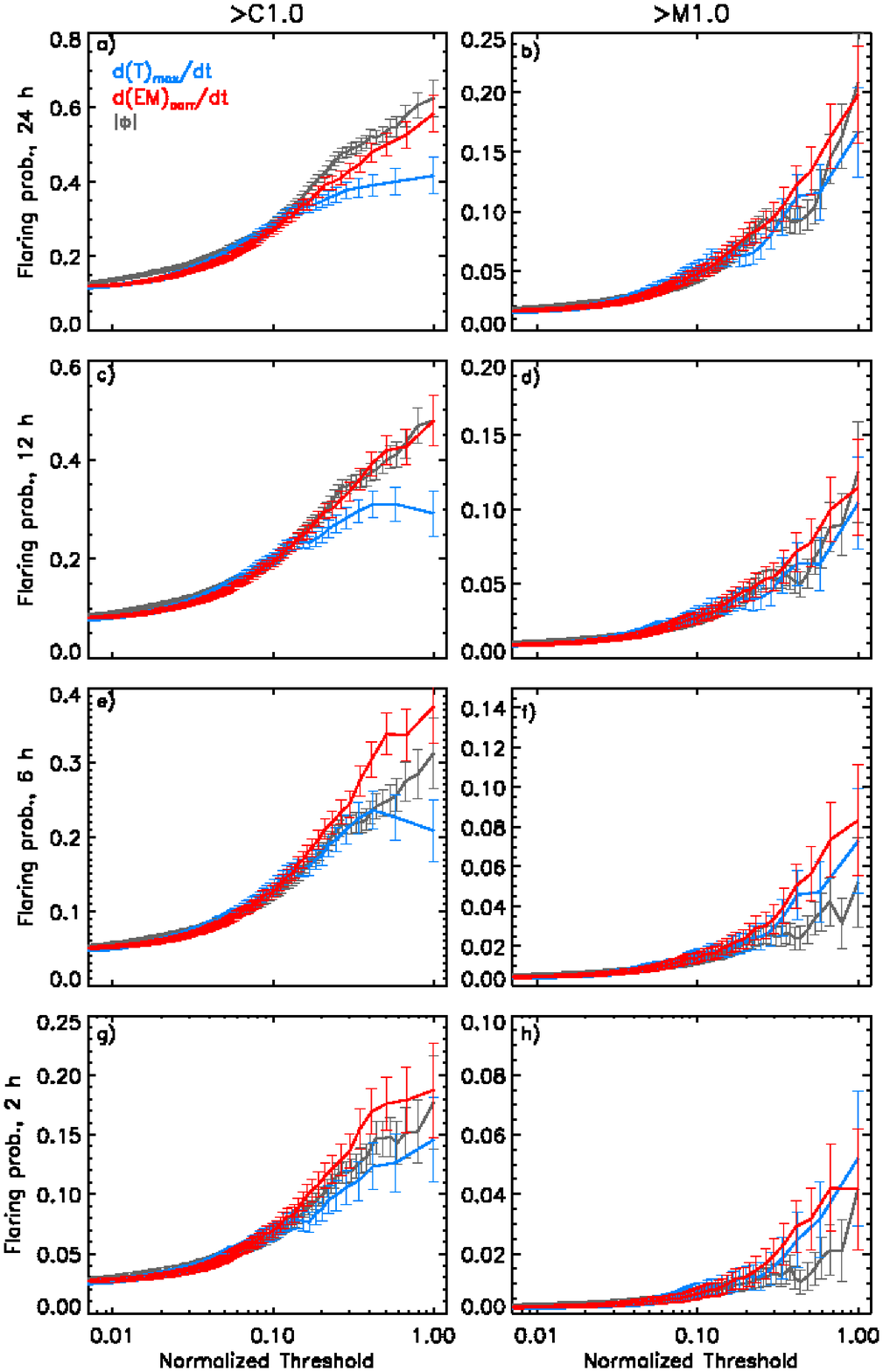}
\caption{Conditional probabilities calculated for a window of flare occurrence of 24, 12, 6 and 2~hours (first, second, third and fourth rows, respectively). The first column of panels shows the probability of the occurrence of a flare higher than C.1, while the second  column, the probability for flares only of M-class and X-class. The probabilities calculated from the $<(d{\rm EM}/dt)^+>$, $<(dT_{\rm max}/dt)^+>$, and $\Phi$ are represented with red, blue and grey  colors respectively. The normalized maxima are $3.\times 10^{27}$~cm$^{-5}/$hr for $<(d{\rm EM}/dt)^+>$, $2.7\times\ 10^6$~K/hr for $<(dT_{\rm max}/dt)^+>$. and $1.6\times 10^8$~Mx for $\Phi$.}\label{cond_probab_fig}
\end{figure*}

Next, we calculated the probability that a solar region will produce a flare within the next 24, 12, 6 and 2\,hours after $t_{\rm SHARP}$, 
on the condition that the DEM-related quantity exceeds given threshold values. We will compare the conditional probabilities derived 
for $< (d{\rm EM}/dt)^+>$ and $< (dT_{\rm max}/dt)^+>$ with those calculated for the unsigned magnetic flux $\Phi$. 
 
The conditional probability $p({\rm th})$ for a threshold (th) is calculated via the equation:
 
\begin{equation}
p({\rm th})\,=\, \frac{N_f+1}{N+2},\label{condproba}
\end{equation}
 
\noindent where $N_{f}$ is the number of flaring regions and $N$ is the total number of regions with parameter value higher than a given threshold \citep{{2005PASA...22..153W},{2012SoPh..276..161G}}. Clearly, $N_f \leq\ N$. 
The error of the probability is given by:
 
\begin{equation}
\delta p({\rm th})\,=\, \sqrt{\frac{ p({\rm th})(1-p({\rm th}))}{N\,+\,3}}.\label{delta_condproba}
\end{equation}
 
We computed the conditional probability $p({\rm th})$ and its error $\delta p({\rm th})$ for an array of thresholds and different forecast windows. 
For each parameter, thresholds were defined by dividing the 9083 values in 100 bins with equal population (roughly 90 data-points each) starting from the maximum value. In order to 
facilitate the comparison between the conditional probabilities of different predictors, $< (d{\rm EM}/dt)^+>$, $< (dT_{\rm max}/dt)^+>$ 
and $\Phi$, the thresholds of each parameter were scaled to their corresponding maximum values, as follows:
  
\begin{equation}
{\rm th}_{{\rm sc}}=\frac{{\rm th}}{{\rm max(th)}}.\label{scale_definition}
\end{equation}
The calculated conditional probabilities are shown in Figure~\ref{cond_probab_fig}.

As a general remark, we can see that the probabilities are overall higher for longer time windows. Probabilities corresponding to the longest forecast window, 24 hours (Figs.~\ref{cond_probab_fig}a,~\ref{cond_probab_fig}b) are higher than the ones corresponding to a 2-hour forecast window 
(Figs.~\ref{cond_probab_fig}g,~\ref{cond_probab_fig}h) by more than a factor of 2. This can be attributed to the larger number of flares occurring 
in the longer time windows and the apparently improved statistics. 
For similar reasons the conditional probabilities are higher when we consider all flares (Figs.~\ref{cond_probab_fig} left column) 
instead of considering only the M- and X-class flares (Figs.~\ref{cond_probab_fig} right column).
 
Figure~\ref{cond_probab_fig} shows that the conditional flaring probabilities computed for $\Phi$ dominate when we consider 
all flares within 24 hours. 
For all flares and time windows equal to 6 and 2\,hours, the $< (d{\rm EM}/dt)^+>$ produces 
higher probabilities, for normalized thresholds higher than 0.3. 
 
Regarding major flares, the $< (d{\rm EM}/dt)^+>$ produces higher probabilities than $\Phi$ for thresholds higher than 0.3, 
for all time windows. However, for shorter time windows, this difference is more pronounced.  Overall the probabilities derived 
for $< (dT_{\rm max}/dt)^+>$ are the lowest, except for major flares in 6 and 2\,hours time windows.
For relative thresholds lower than 0.2, the probability functions are similar between the three parameters and therefore, the $< (d{\rm EM}/dt)^+>$ and $< (dT_{\rm max}/dt)^+>$ has the same ability to forecast flare-quiet areas as $\Phi$ but maybe reinforced for cases with negative slope that were not studied in this work.

In conclusion, for short time windows, $< (d{\rm EM}/dt)^+>$ yields higher flaring conditional probabilities 
than the baseline unsigned flux for the 2-hr  and the 6-hr windows respectively. This finding indicates that the 
time evolution of EM is potentially useful as a short term forecasting tool.

\section{Discussion and Conclusions}
In this work we explored whether the DEM temporal evolution can be used as a possible short-term 
precursor or predictor of solar flares. The choice of the DEM parameters was inspired by the results of \citet{2016A&A...588A..16S}, 
who found a rise of the DEM hot components before two X-class flares. DEM is a measure of the heating of the EUV emitting 
plasma. Therefore monitoring the DEM  as a function of time can indicate potential pre-heating activity in preflare stages.
DEM is also capable to disentangle the plasma emission at various temperatures which are mixed together in the SDO/AIA images. 
A large number of cases is necessary to verify how common are the findings of \citet{2016A&A...588A..16S}. For this reason 
we used the GAIA-DEM database provided by the Institut d'Astrophysique Spatiale in Orsay, France, which includes a large 
collection of DEM maps, derived from SDO/AIA coronal images. This database provides an easy access to a large statistical 
sample of active regions, allowing the study of their evolution. 
 
We examine whether preflare heating can be monitored by a preflare increase of EM, $T_{\rm max}$ and $\sigma$, 
the three Gaussian parameters that describe the DEM distributions. The decrease of these parameters corresponds to plasma 
cooling, which can happen, for example after a flare. For this reason we constrained our study to heating by considering only 
the positive values of the DEM parameter derivatives from the timeseries. Plasma heating manifests itself continuously, often 
in flares occurring before a major flare event. It is known that the occurrence of flares or flashes in a given active region is indicative 
of imminent flares  \citep{2012ApJ...757...32F,2005PASA...22..153W,1991SoPh..131..149Z}. In our study we tried to differentiate between cases 
of gradual rising of timeseries and of sudden energy release through flares prior to the main event in order to investigate the relative importance 
of the two types of evolution in flare prediction. In both cases, care was taken not to include in our measurements the onset of the flare intending to predict. 

The derivatives where calculated using a SG filtering in order to smooth the peaks in the timeseries and to enhance the effect of the continuous heating events. We selected a three degree polynomial for the smoothing procedure but we also tested 2-degree and 5-degree polynomials and we also calculated derivatives using central-differences. For all these cases we found qualitatively similar results.

The study of 16 active regions indicated that the positive values of the derivatives $(d{\rm EM}/dt)^+$ and $(dT_{\rm max}/dt)^+$ 
parameters are more sensitive to an imminent flare. The $(d{\rm EM}/dt)^+$ and $(dT_{\rm max}/dt)^+$ not associated with 
future flares have, statistically, lower values than the values preceding flares. Unfortunately, this tendency is not without overlapping, 
which was very pronounced in the  $(d\sigma/dt)^+$ distributions. As a result, $(d\sigma/dt)^+$ was not further considered.
 
Further motivated by these preliminary results, we analysed a statistically significant sample of point-in-time observations 
and we investigated the efficiency of $< (d{\rm EM}/dt)^+>$ and $< (dT_{\rm max}/dt)^+>$ as flare predictors. 
This was tested for four time windows, namely, 2, 6, 12 and 24\,hours,
 
The probabilities derived from the DEM parameters are higher, than the ones derived from the unsigned magnetic flux, 
for major flares, and shorter time windows for scaled threshold values higher than 0.3. Overall, the $< (d{\rm EM}/dt)^+>$ 
yields higher probabilities than $< (dT_{\rm max}/dt)^+>$.

This result shows that the EM is more sensitive to pre-flare heating than $T_{max}$. Perhaps through these 
heating events the DEM distribution widens towards higher temperatures. The simplistic approach of a Gaussian 
fit may, then, not be able to locate changes in the position of the maximum ($T_{max}$) but can fit the integral over the distribution (EM). 
 
Point taken, there are some caveats inherent to the present analysis. The use of a Gaussian as a function of the logarithmic 
temperature, is only a first-order description of the DEM function and does not provide a direct diagnostic of the temporal evolution 
of the hot, 7-12~MK plasma involved in the preflare state. Also, the timeseries shown are computed from the generally  
large SHARP regions, often exceeding typical active-region sizes.  Moreover, active regions were treated as a whole and we did not focus on 
specific morphologies of active regions, such as polarity inversion lines, or parasitic polarities that are expected to play an important 
role in the production of flares \citep{2007ApJ...655L.117S, 2007ApJ...661L.109G, 2002ApJ...569.1016F, 2006ApJ...644.1258F, Kontogiannis17}. 
 
In any case, our results imply that the DEM parameters may, in principle, serve as a precursor or even predictor of solar flares, 
with success greater than that of the unsigned magnetic flux. In the future, more sophisticated analyses, based on better 
refined DEM function and restricted to certain features of active regions could lead an enhanced flare forecasting capability. 

{\bf Acknowledgements}
This research has been funded by the European Union's Horizon2020 research and innovation programme Flare Likelihood And Region Eruption foreCASTing (FLARECAST) project, under grant agreement N0. 640216. 
This work used data provided by the GAIA-DEM archive. GAIA-DEM (http://medoc-dem.ias.u-psud.fr/) is a service provided by the MEDOC data and operations centre (CNES / CNRS / Univ. Paris-Sud), http://medoc.ias.u-psud.fr. 
AIA data are courtesy of the NASA/SDO and AIA science teams. P. Syntelis received funding from the Science and Technology Facilities Council (UK) through the consolidated grant ST/N000609/1.

\textbf{Disclosure of Potential Conflicts of Interest.} The authors 
declare that they have no conflicts of interest.

\bibliography{bibliographyflar2}

\end{document}